\documentclass[10pt,onecolumn,twoside]{article}

\pdfoutput=1

\usepackage{amssymb}
\usepackage{amsmath}
\usepackage{graphicx}
\usepackage{bm}
\usepackage{mathtools}
\usepackage{multirow}
\usepackage{url}
\usepackage{fullpage}
\usepackage{cite}
\usepackage{authblk}

\def\R{\mathbb{R}}
\def\Z{\mathbb{Z}}
\def\epsbf{\bm{\epsilon}}

\def\phibf{\bm{\phi}}
\def\phibfhat{\widehat{\bm{\phi}}}
\def\psibf{\bm{\psi}}
\def\varphibf{\bm{\varphi}}
\def\kbf{\bm{k}}
\def\wbf{\bm{w}}
\def\sbf{\bm{s}}
\def\dbf{\bm{d}}
\def\xbf{\bm{x}}
\def\ybf{\bm{y}}
\def\zbf{\bm{z}}
\def\ebf{\bm{e}}
\def\wrap{\mathcal{W}}
\def\Dbf{\mathbf{D}}
\def\Hbf{\mathbf{H}}
\def\defn{\,\triangleq\,} 

\def\infbf{\bm{\infty}}

\def\shatten{\mathcal{S}_1}

\def\Phibf{\bm{\Phi}}

\def\dataterm{\mathcal{D}}
\def\regularizer{\mathcal{R}}
\def\lagrangian{\mathcal{L}}
\def\shrinkage{\mathcal{T}}

\def\argmin{\mathop{\mathrm{arg\,min}}}

\begin{document}

\title{Isotropic inverse-problem approach  for two-dimensional phase unwrapping}

\author[1,2]{Ulugbek~S.~Kamilov
\thanks{The work of U.~S.~Kamilov and M.~Unser was
supported by the European Research Council under the
European Union's Seventh Framework Programme
(FP7/2007-2013)/ERC Grant Agreement 267439.
This work was conducted when U.~S.~Kamilov was
with Biomedical Imaging Group, {\'{E}}cole polytechnique
f{\'{e}}d{\'{e}}rale de Lausanne, Switzerland. He is now with
Mitsubishi Electric Research Laboratories, Cambridge,
MA, USA. Corresponding author: kamilov@merl.com.
}}
\author[3]{Ioannis~N.~Papadopoulos}
\author[3]{Morteza~H.~Shoreh}
\author[3]{Demetri Psaltis}
\author[1]{Michael Unser}
\affil[1]{Biomedical Imaging Group, {\'{E}}cole polytechnique f{\'{e}}d{\'{e}}rale de Lausanne, Switzerland}
\affil[2]{Mitsubishi Electric Research Laboratories, Cambridge, MA 02139, USA}
\affil[3]{Optics Laboratory, {\'{E}}cole polytechnique f{\'{e}}d{\'{e}}rale de Lausanne, Switzerland}

\maketitle 

\begin{abstract}In this paper, we propose a new technique for 
two-dimensional phase unwrapping. The unwrapped 
phase is found as the solution of an inverse 
problem that consists in the minimization of 
an energy functional. The latter includes 
a weighted data-fidelity term that favors sparsity in 
the error between the true and wrapped phase differences,
as well as a regularizer based on higher-order 
total-variation. One desirable feature of our method 
is its rotation invariance, 
which allows it to unwrap 
a much larger class of images 
compared to the state of the art.
We demonstrate 
the effectiveness of our method 
through several experiments on simulated and real data
obtained through the tomographic phase microscope.
The proposed method can enhance the applicability and 
outreach of techniques that rely on quantitative phase 
evaluation.
\end{abstract}

\tableofcontents

\newpage


\section{Introduction}

Two-dimensional phase unwrapping is an essential component 
in a majority of techniques used for quantitative phase imaging. 
A typical example of standard
application is tomographic phase microscopy~\cite{Choi.etal2007}, 
where the phase of 
the transmitted wave-field must be first unwrapped before being 
interpreted as a line integral of the refractive index along 
the direction of propagation. Phase unwrapping 
is also used for estimation of terrain elevation 
in synthetic aperture radar~\cite{Rosen.etal2000}, 
wave/fat  separation in magnetic resonance 
imaging~\cite{Song.etal1995}, and estimation of
wave-front distortion in adaptive optics~\cite{Fried1977}.
Accordingly, improvements in phase unwrapping 
methodology can enhance the applicability and 
outreach of techniques that rely on quantitative phase 
evaluation.

The centrality of phase unwrapping has 
resulted in the development of many practical solutions to 
the problem. Such solutions include direct approaches
based on path-following such as the two-dimensional extension 
of the well known Itoh's method~\cite{Itoh1982} as well as 
more evolved strategies based on branch 
cuts~\cite{Goldstein.etal1988} or quality maps~\cite{Herraez.etal2002}.
The current trend in the literature is to formulate the task of
phase unwrapping as an inverse problem in a path-independent way. 
Earlier works have proposed to minimize 
the quadratic error between the true and wrapped 
phase differences~\cite{Hunt1979, Takajo.Takahashi1988, Ghiglia.Romero1994}. 
Marroquin and Rivera~\cite{Marroquin.Rivera1995}  have applied 
Tikhonov regularization 
to improve and stabilize the performance of the 
least-squares approach. 
They showed that the introduction of the regularization 
term permits the algorithm to cope with noise and 
missing data. Huang \textit{et al.}~\cite{Huang.etal2012}
showed that the performance of phase unwrapping 
is further improved when replacing Tikhonov with 
total-variation regularization.
In the context of phase unwrapping,
Ghiglia and Romero~\cite{Ghiglia.Romero1996}
recognized the tendency of quadratic-penalty 
to smooth image edges and proposed to minimize a more
general $\ell_p$ norm--based criterion as an alternative.
They found that the performance of phase
unwrapping improves when $0 \leq p \leq 1$, albeit at the 
increase in computational cost. Similarly, 
Rivera and Marroquin~\cite{Rivera.Marroquin2004} have investigated nonconvex
optimization strategies relying on half-quadratic regularization. 
More recently, Gonz{\'{a}}lez and Jacques~\cite{Gonzalez.Jacques2014}
have proposed an iterative unwrapping method based on
$\ell_1$ minimization that additionally promotes 
sparsity of the solution in the wavelet domain.
Ying \textit{et al.}~\cite{Ying.etal2006} 
have proposed an iterative method based on dynamic programming
that models the phase as a Markov random field.
Bioucas-Dias and 
Valad{\~{a}}o~\cite{Bioucas-Dias.Valadao2007}
have proposed an energy functional based on generalized $\ell_p$ norm
and corresponding minimization algorithm that relies on graph-cut 
methods. Their PUMA algorithm in its original form and its 
noise tolerant 
extensions~\cite{Bioucas-Dias.etal2008, Valadao.Biocas-Dias2009} 
are currently considered state of the art. 
Mei \textit{et al.}~\cite{Mei.etal2013} 
have proposed an application specific method 
that jointly unwraps and denoises time-of-flight phase images 
using a message-passing algorithm. 
An extended review of this topic, along with related 
algorithmic ideas, can be found in the book~\cite{Ghiglia.Pritt1998} 
and tutorial~\cite{Ying2006}.

In this paper, we propose a new variational-reconstruction approach
for phase unwrapping that is robust to noise. In particular, our aim is to improve on 
state of the art by introducing an improved energy 
functional and demonstrating its benefits. The main
contributions of this paper can be summarized as follows:
\begin{itemize}

\item Formulation of the unwrapping as an optimization problem
where the data-fidelity term penalizes the weighted $\ell_1$-norm
of the error in a way that is invariant to rotations. Our formulation 
thus allows the phase image to contain edges that are of 
arbitrary orientation, which is
distinct from traditional approaches in literature~\cite{Ghiglia.Romero1996, Rivera.Marroquin2004, 
Ying.etal2006, Bioucas-Dias.Valadao2007, Gonzalez.Jacques2014}.

\item Use of a non-quadratic regularization term that allows our method
to cope with noise, while still preserving sharp edges in the phase image.
Our regularizer consists of a higher-order extension of total-variation (TV)
that is currently considered state of the art in the context of 
resolution of linear inverse problems in biomedical imaging~\cite{Lefkimmiatis.etal2013}.

\item Design of a novel iterative algorithm for phase unwrapping.
The algorithm approximates the minimum $\ell_0$-norm solution
by solving a sequence of weighted
$\ell_1$-norm minimization problems, where the weights at the 
next iteration are computed from the value of the current solution.
Since our energy functional is non-smooth, we rely 
on a well known alternating direction method of multipliers 
(ADMM)~\cite{Boyd.etal2011} to decompose the 
minimization into a sequence of simpler operations.

\end{itemize}

This paper is organized as follows. In Section~\ref{Sec:Problem}, we 
introduce our formulation of the phase unwrapping, and discuss 
the relevance of this new approach for obtaining high-quality 
solutions in practice. In Section~\ref{Sec:Algorithm}, we derive 
our reconstruction algorithm. In 
Section~\ref{Sec:Experiments}, we conduct experiments on 
simulated and real phase unwrapping problems, and compare 
our method with the state of the art from both 
qualitative and quantitative standpoints. We summarize and
conclude our work in Section~\ref{Sec:Conclusion}.

\section{Problem formulation} \label{Sec:Problem}

We consider the following observation model
\begin{equation} \label{Eq:Model}
\psibf = \wrap\left(\phibf\right) = \phibf - 2\pi\kbf
\end{equation}
where $\kbf \in \Z^N$ and $\phibf, \psibf \in \R^N$. The vectors 
$\psibf$ and $\phibf$ 
represent vectorized versions of the wrapped and unwrapped 
phase images, respectively.
The wrapping is represented by a component-wise 
function $\wrap$ that is defined as
\begin{equation*}
\wrap(\phi) \defn \left[\left(\left(\phi + \pi\right) \hspace{-0.5em}\mod 2\pi\right) - \pi\right] \quad\in\quad \left[-\pi, \pi\right).
\end{equation*}
When noise is part of the measurements, we assume that the unwrapped 
phase vector $\phibf$ in our model represents the noisy version of the true 
phase $\varphibf$.

Generally, the two-dimensional phase unwrapping problem 
is ill-posed. However, it can be solved exactly in the 
noiseless scenario, when the phase $\phibf$ satisfies 
the two-dimensional extension of Itoh's continuity condition~\cite{Itoh1982}. 
Let $\Dbf : \R^N \rightarrow \R^{N \times 2}$ denote the 
discrete counterpart of the gradient operator and let
\begin{equation}
\Dbf\phibf = 
\begin{bmatrix} 
\Dbf_x \phibf \\[0.3em]
\Dbf_y \phibf
\end{bmatrix},
\end{equation}
where $\Dbf_x$ and $\Dbf_y$ denote the finite-difference operator 
along the horizontal and vertical directions, respectively. If, for a given 
pixel $n \in [1, \dots, N]$, the unwrapped phase $\phibf$ satisfies
\begin{equation} \label{Eq:Itoh}
\|[\Dbf\phibf]_n\|_{\ell_2} \defn \sqrt{([\Dbf_x\phibf]_n)^2+([\Dbf_y\phibf]_n)^2}
\leq \pi,
\end{equation}
then, we have the equality
\begin{equation} \label{Eq:GradEquality}
[\Dbf\phibf]_n = \wrap([\Dbf\psibf]_n).
\end{equation}
Here, $[\Dbf\phibf]_n \defn ([\Dbf_x\phibf]_n, [\Dbf_y\phibf]_n)$ denotes the $n$-th component 
of the gradient $\Dbf\phibf$. Relation~\eqref{Eq:GradEquality} 
suggests that two-dimensional phase unwrapping
may be accomplished by a simple phase summation, provided 
that~\eqref{Eq:Itoh} is satisfied at all pixels $n$. Note that
the formulation in~\eqref{Eq:Itoh} imposes 
the Itoh's continuity condition on both gradient components 
simultaneously due to the norm inequality
\begin{equation*}
\max\left(\left|a\right|, \left|b \right|\right)
\leq \sqrt{a^2+b^2},
\end{equation*}
which holds for any $a, b \in \R$.

In practice, however, condition~\eqref{Eq:Itoh} will not be
fulfilled at all pixel locations due to the presence of sharp
edges and of measurement noise. 
Yet, it can still be expected to hold for the great majority 
of pixels of the 
unwrapped phase image. We thus formulate phase unwrapping as the
following minimization problem
\begin{equation} \label{Eq:VariationalProblem}
\phibfhat = \argmin_{\phibf \in \Phibf} \left\{\dataterm(\phibf) + \tau \regularizer(\phibf) \right\},
\end{equation}
where $\dataterm$ is the data-fidelity term and $\regularizer$ is the regularization term,
to be discussed shortly. The convex set 
$\Phibf = \{\phibf \in \R^N \,:\, \phi_1 = \psi_1\} \subset \R^N$ enforces the 
first pixel of the solution $\phibf$
to match the first pixel of the wrapped phase $\psibf$, which removes 
the additive constant ambiguity present in phase unwrapping. The parameter
$\tau > 0$ controls the amount of regularization.

The data-fidelity term in~\eqref{Eq:VariationalProblem} is given by
\begin{equation}
\label{Eq:DataTerm}
\dataterm(\phibf) 
\defn \sum_{n = 1}^N w_n \|[\Dbf\phibf-\wrap(\Dbf\psibf)]_n\|_{\ell_2},
\end{equation}
where $\wbf \in \R_+^N$ are positive weights. It is intended to relax 
the strict equality~\eqref{Eq:GradEquality}.

In the unweighted case,
i.e.\ when $w_n = 1$ for all $n \in [1,\dots,N]$, 
$\dataterm$ corresponds to
$\ell_1$-norm penalty on the magnitudes of $\epsbf \defn \Dbf\phibf-\wrap(\Dbf\psibf)$.
It can be interpreted as a convex relaxation of $\ell_0$-norm 
penalty that enforces sparse magnitudes of $\epsbf$. 
This implies that our data-term favors $\phibf$ whose gradient $\Dbf\phibf$
agrees with $\wrap(\Dbf\psibf)$ on most of the pixels.
Moreover, our
data-fidelity term $\dataterm$ penalizes both horizontal $\epsbf_x$ 
and vertical $\epsbf_y$ components of $\epsbf$ in a joint fashion. 
This is significantly
different from traditional formulations in 
the literature~\cite{Ghiglia.Romero1996, Gonzalez.Jacques2014},
where the $\ell_1$-norm is penalized in a separable fashion 
as $\|\epsbf\|_{\ell_1} = \|\epsbf_x\|_{\ell_1} + \|\epsbf_y\|_{\ell_1}$. In fact,
there is a clear analogy between our formulation~\eqref{Eq:DataTerm}
and the \textit{isotropic}, i.e.\ rotation invariant, form of total-variation (TV) that is 
often used for edge-preserving image restoration~\cite{Beck.Teboulle2009a}.
Similarly, the separable $\ell_1$-penalty $\|\epsbf\|_{\ell_1}$ is analogous to
the \textit{anisotropic} form of TV. The arbitrary 
orientation of edges in a typical image makes isotropic 
TV penalty a preferred choice for image restoration. 
The numerical
experiments presented in Sec.~\ref{Sec:Experiments} illustrate that
indeed our method based on isotropic formulation can unwrap a 
larger class of images compared
to other state-of-the-art phase unwrapping methods.

It has been reported in several works~\cite{Ghiglia.Romero1996, Bioucas-Dias.Valadao2007} that nonconvex approaches 
based on $\ell_p$-norm penalization
of $\epsbf$, with $0 \leq p < 1$, further improve the performance
of phase unwrapping.  In particular, $\ell_0$-norm is generally 
accepted as the most desirable in practice. One of the properties
of $\ell_1$-norm that distinguishes it from $\ell_0$ is that 
it takes into account 
the actual values of the magnitudes of $\epsbf$, whereas $\ell_0$-norm
disregards this information and only counts the support. One possible strategy for selecting  
weights in the context of $\ell_1$ minimization 
proposed by Cand{\`e}s \textit{et al.}~\cite{Candes.etal2008} is to pick $\wbf$
such that it counteracts the influence of the magnitude on the
$\ell_1$-norm. For example, suppose that the
weights were inversely proportional to the true magnitudes
\begin{equation} \label{Eq:IdealWeights}
w_n =
\begin{cases}
\frac{1}{\|[\epsbf]_n\|_{\ell_2}} & \text{if } \|[\epsbf]_n\|_{\ell_2} \neq 0 \\
\infty       & \text{if } \|[\epsbf]_n\|_{\ell_2} = 0.
\end{cases}
\end{equation}
Then, if there are exactly $m$ pixels violating~\eqref{Eq:Itoh},
the minimizer in~\eqref{Eq:VariationalProblem} is guaranteed to find the solution
corresponding to $\ell_0$ data-fidelity term. 
The large values in $\wbf$ force the
solution $\phibfhat$ to concentrate on the pixels where the weights
are small, and by construction these correspond precisely to the 
pixels where the magnitudes of $\epsbf$ are nonzero. Typically,
these pixels correspond to the area of the phase image that
contains a sharp edge. It is clearly
impossible to construct the precise weights~\eqref{Eq:IdealWeights}
without knowing the unwrapped phase $\phibf$ itself,
but this suggests more generally that large weights could be used 
to discourage nonzero magnitudes in $\epsbf$, 
while small weights could be used to encourage nonzero 
magnitudes in $\epsbf$.

\begin{figure*}[t]
\centering\includegraphics[width=13cm]{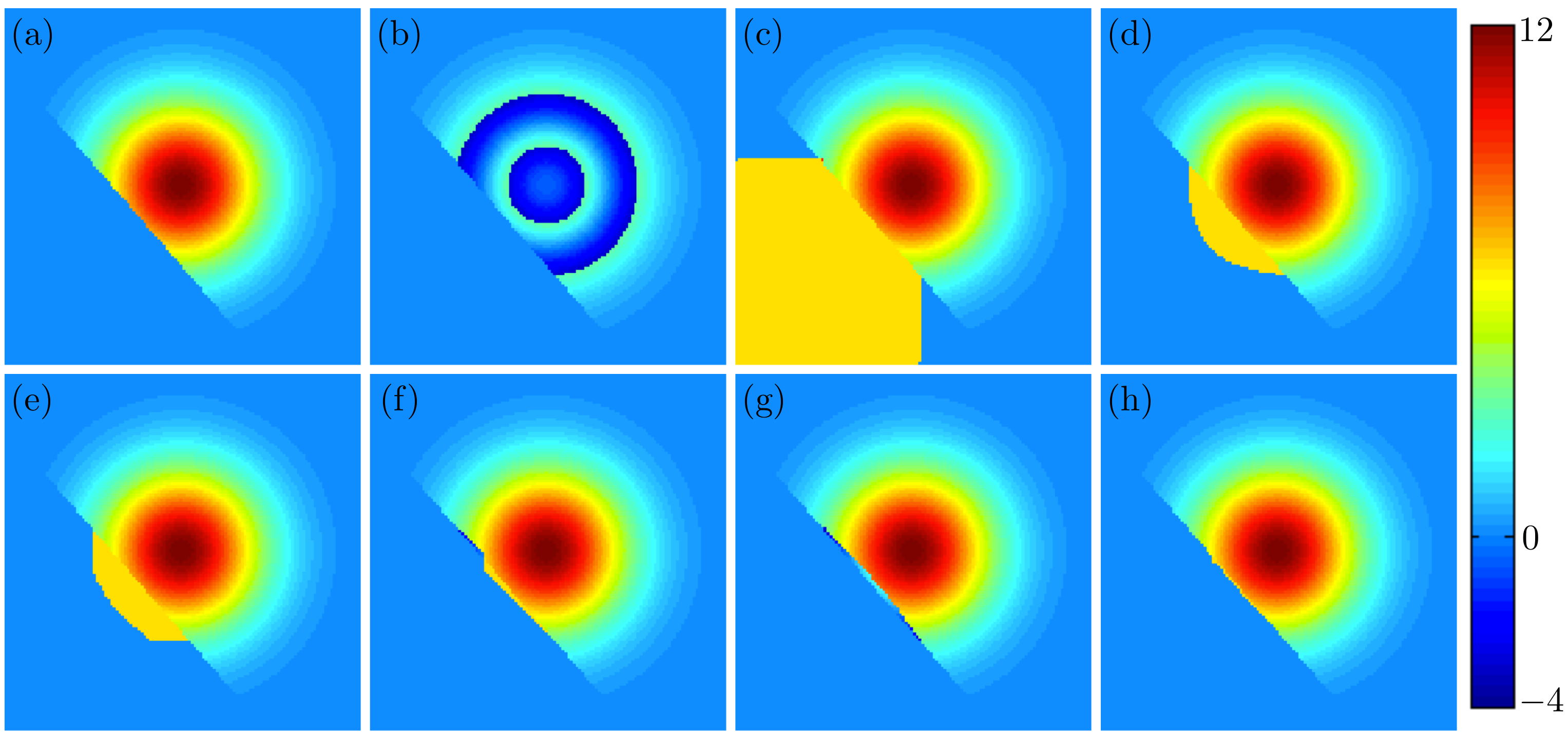}
\caption{Phase unwrapping of $128 \times 128$ image
of a truncated Gaussian function. (a) true phase, (b) wrapped phase, 
(c) Goldstein's algorithm (GA), (d) least-squares (LS),
(e) iteratively reweighted LS (IRLS), (f) PUMA,
(g) proposed method with uniform weights, 
(h) proposed method with adaptive weights (IRTV).}
\label{Fig:Gaussian}
\end{figure*}

As regularization term in~\eqref{Eq:VariationalProblem},
we propose to use the Schatten norms 
of the Hessian matrix at every pixel of the 
image~\cite{Lefkimmiatis.etal2013}. Specifically, we set
\begin{subequations}
\begin{align}
\regularizer(\phibf) 
&\defn \|\Hbf\phibf\|_{1, \shatten} \\
&\defn \sum_{n = 1}^N (\sigma_{1}([\Hbf\phibf]_n) + \sigma_{2}([\Hbf\phibf]_n)),
\end{align}
\end{subequations}
where $\Hbf : \R^N \rightarrow \R^{N \times 2 \times 2}$ is the 
discrete Hessian operator and $\|\Hbf\phibf\|_{1, \shatten}$ is
the nuclear norm that is 
computed by summing singular values $\sigma_{1}([\Hbf\phibf]_n)$
and $\sigma_{2}([\Hbf\phibf]_n)$ of the Hessian matrix 
at position $n$. There are four major advantages of using 
such Hessian Schatten-norm (HS) regularization: 

\begin{itemize}

\item As a higher order regularizer HS avoids the staircase effect of
TV and results in piecewise-smooth variations of intensity in 
the reconstructed phase image. Accordingly, this makes HS
particularly well suited for biological and medical specimens that 
consist of complicated structures such as filaments.

\item Similarly to the data-fidelity term $\dataterm$, 
our regularizer $\regularizer$
is convex and rotation invariant~\cite{Lefkimmiatis.etal2013}. 
Rotation invariance implies that we can expect it to work equally 
well on phase images that contain objects of arbitrary orientations.

\item HS regularization has been shown to be state of the art 
in resolution of linear inverse problems. In particular, it 
was shown in~\cite{Lefkimmiatis.etal2013} that HS consistently 
outperforms other popular regularizers such as Tikhonov, wavelet, and 
TV.

\item Convexity and its algebraic structure make HS 
amenable to efficient algorithmic implementation 
and thus practical for large 
scale inverse problems that are typical in imaging.

\end{itemize}

To demonstrate the performance of our variational approach~\eqref{Eq:VariationalProblem}, 
we present a phase unwrapping
experiment on a synthetic
image consisting of a 2D Gaussian function of amplitude $12$
and standard deviation $20$ that has been truncated along 
a line of arbitrary orientation (here about $40^\textrm{o}$). 
Fig.~\ref{Fig:Gaussian} illustrates the results for
four standard unwrapping methods such as Goldstein's algorithm (GA)~\cite{Goldstein.etal1988},
least-squares (LS)~\cite{Ghiglia.Romero1994},
iteratively reweighted LS (IRLS) 
with data-dependent weights that
approximate the $\ell_0$-norm penalty~\cite{Ghiglia.Romero1996},
and PUMA~\cite{Bioucas-Dias.Valadao2007}. We additionally illustrate
the performance of our method with the emphasis on the influence
of weights $\wbf$ over the final solution $\phibfhat$. 
Accordingly, we show the solution of~\eqref{Eq:VariationalProblem} 
with uniform and data-dependent
weights. 
As expected, all algorithms perform equally well in the continuous
region of the image. On the other hand, our approach is the only 
one that accurately captures the discontinuous region of the unwrapped 
image. This is expected due to rotation invariance of our energy
functional. Additionally, we note that the LS method, which is also based on
rotation invariant energy functional, fails to preserve the 
edge due to excessive smoothing. Finally, a careful inspection of 
Figs.~\ref{Fig:Gaussian} (g) and (h) reveals that the edge is much
sharper when the weights $\wbf$ are selected in a data-dependent fashion as
explained next in Sec.~\ref{Sec:Algorithm}.

\section{Reconstruction algorithm} \label{Sec:Algorithm}

We now describe our computational approach based on
the convex optimization 
problem~\eqref{Eq:VariationalProblem}. The
iterative scheme alternates between
estimating $\phibf$ and redefining the weights $\wbf$
as follows.
\begin{enumerate}

\item \textit{Initialization: } Set iteration number to $t = 1$. Select an initial phase $\phibf^0 \in \R^N$ and 
set $w_n^0 = 1$ for each $n = 1, \dots, N$. 

\item \textit{Optimization: } For a fixed $\wbf^{t-1}$, compute the phase image $\phibf^t$ by solving~\eqref{Eq:VariationalProblem}. Also, compute the auxiliary variable $\epsbf^t = \Dbf\phibf^t-\wrap(\Dbf\psibf)$.

\item \textit{Weight adaptation: } For each $n = 1, \dots, N$
\begin{equation}
w_n^{t} =
\begin{cases}
\frac{1}{\|[\epsbf^t]_n\|_{\ell_2}} & \text{if } \epsilon_{\textrm{min}} \leq \|[\epsbf^t]_n\|_{\ell_2} \leq \epsilon_{\textrm{max}} \\
\frac{1}{\epsilon_{\textrm{max}}}       & \text{if } \|[\epsbf^t]_n\|_{\ell_2} \geq \epsilon_{\textrm{max}} \\
\frac{1}{\epsilon_{\textrm{min}}}       & \text{if } \|[\epsbf^t]_n\|_{\ell_2} \leq \epsilon_{\textrm{min}} \\
\end{cases}
\end{equation}

\item Stop on convergence or when $t$ attains a specific maximum number of iterations $t_{\textrm{max}}$. 
Otherwise, increment $t$ and proceed to step 2.

\end{enumerate}
The two parameters $\epsilon_{\textrm{min}}$ and $\epsilon_{\textrm{max}}$ in step 3 
provide stability and avoid divisions by zero. For our experiments in Sec~\ref{Sec:Experiments}, we set $\epsilon_{\textrm{max}} = 1/\epsilon_{\textrm{min}} = 10$. 
The optimization in step 2 will be discussed shortly.

Although the initial phase $\phibf^0$
can be set to an arbitrary vector in $\R^N$, in practice, a warm initialization leads
to a smaller number of iterations required for convergence, and hence
faster unwrapping times. In our experiments,
we found that the solution of LS can serve as a computationally inexpensive way of obtaining
a good initialization.

Using an adaptive approach to construct the weights progressively
improves the unwrapping
around the discontinuities in the phase image. These phase discontinuities 
might be due to a presence of a sharp edge 
or due to strong noise. Even though the early phase estimate may be inaccurate,
the largest coefficients of $\epsbf$ are most likely to be identified with a phase discontinuity. 
Once these locations are identified, their influence is downweighted 
in order to gain in sensitivity for identifying the remaining regions of the phase image.

\begin{figure*}[t]
\centering\includegraphics[width=16.5cm]{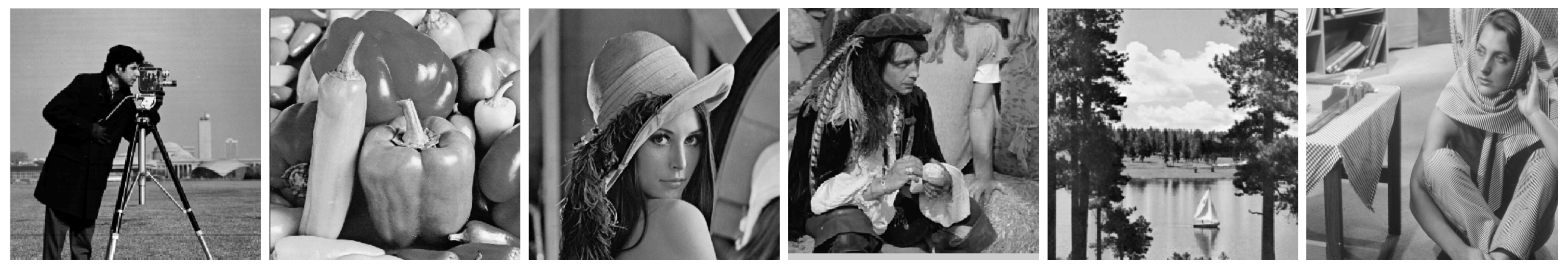}
\caption{Set of standard test images. 
From left to right: Cameraman, Peppers, Lena, Man, Lake, and Barbara.}
\label{Fig:ImageDatabase}
\end{figure*}

\begin{table*}[t]
\begin{center}
  \caption{Performance of the five considered phase 
  unwrapping methods in terms of SNR on the set of
  standard test images for various amplitudes}\label{Tab:Comparison}
\end{center}
\begin{minipage}[b]{0.5\textwidth}
  \scriptsize
  	\bgroup
	\def\arraystretch{1.25}
	\setlength{\tabcolsep}{1em}
	
    \begin{tabular}{|c c || c | c | c | c  | c | }
    \hline
    \multicolumn{2}{ |c ||  }{\multirow{2}{*}{Img/Amp} } &
    \multicolumn{5}{ c| }{SNR (dB)}\\ \cline{3-7}
    \multicolumn{2}{ |c||  }{} & GA & LS & IRLS & PUMA & IRTV \\

    \hline\hline
    \multicolumn{1}{|c|}
    {\multirow{6}{*}{\rotatebox[origin=c]{90}{\textbf{Cameraman}}}} & 
    4 & 33.99 & 38.76 & 32.74 & 35.72 & $\infbf$ \\ \cline{2-7}
    
    \multicolumn{1}{|c|}{} & 
    5 & 23.38 & \textbf{25.79} & 24.07 & 23.89 & 25.65 \\ 
    \cline{2-7}
    
    \multicolumn{1}{|c|}{} & 
    6 & 14.10 & 15.51 & 19.06 & 17.41 & \textbf{19.98} \\ 
    \cline{2-7}
    
    \multicolumn{1}{|c|}{} & 
    7 & -1.31 & 7.57 & 15.64 & 12.73 & \textbf{16.09} \\ 
    \cline{2-7}
    
    \multicolumn{1}{|c|}{} & 
    8 & 0.00 & \textbf{2.35} & 0.38 & 0.73 & 0.92 \\ 
    \cline{2-7}
    
    \multicolumn{1}{|c|}{} & 
    9 & 1.63 & \textbf{2.64} & 2.02 & 1.90 & 2.17 \\ 
    \cline{1-7}
    
    
    \hline\hline
    \multicolumn{1}{|c|}
    {\multirow{6}{*}{\rotatebox[origin=c]{90}{\textbf{Peppers}}}} & 
    4 & $\infbf$ & $\infbf$ & $\infbf$ & $\infbf$ & $\infbf$ \\ 
    \cline{2-7}
    
    \multicolumn{1}{|c|}{} & 
    5 & 36.69 & $\infbf$ & $\infbf$ & $\infbf$ & $\infbf$ \\ 
    \cline{2-7}
    
    \multicolumn{1}{|c|}{} & 
    6 & 23.64 & 22.60 & $\infbf$ & $\infbf$ & $\infbf$ \\ 
    \cline{2-7}
    
    \multicolumn{1}{|c|}{} & 
    7 & 14.40 & 11.67 & $\infbf$ & 26.39 & $\infbf$ \\ 
    \cline{2-7}
    
    \multicolumn{1}{|c|}{} & 
    8 & 7.27 & 8.23 & 23.96 & 14.93 & \textbf{27.62} \\ 
    \cline{2-7}
    
    \multicolumn{1}{|c|}{} & 
    9 & 5.58 & 3.97 & 15.69 & 15.64 & \textbf{15.75} \\ 
    \cline{1-7}
    
    
    \hline\hline
    \multicolumn{1}{|c|}
    {\multirow{6}{*}{\rotatebox[origin=c]{90}{\textbf{Lena}}}} & 
    4 & $\infbf$ & $\infbf$ & $\infbf$ & $\infbf$ & $\infbf$ \\ 
    \cline{2-7}
    
    \multicolumn{1}{|c|}{} & 
    5 & $\bm{\infty}$ & $\infbf$ & $\infbf$ & $\infbf$ & $\infbf$ \\ 
    \cline{2-7}
    
    \multicolumn{1}{|c|}{} & 
    6 & 38.56 & 27.77 & $\infbf$ & $\infbf$ & $\infbf$ \\ 
    \cline{2-7}
    
    \multicolumn{1}{|c|}{} & 
    7 & 23.62 & 18.70 & $\infbf$ & $\infbf$ & $\infbf$ \\ 
    \cline{2-7}
    
    \multicolumn{1}{|c|}{} & 
    8 & 12.20 & 13.33 & 36.29 & 30.30 & $\infbf$ \\ 
    \cline{2-7}
    
    \multicolumn{1}{|c|}{} & 
    9 & -2.97 & 7.33 & 25.50 & 15.51 & \textbf{30.47} \\ 
    \cline{1-7}
    
    \end{tabular}
    \egroup
\end{minipage}
\begin{minipage}[b]{0.5\textwidth}
	\scriptsize
  	\bgroup
	
	\def\arraystretch{1.25}
	\setlength{\tabcolsep}{1em}
	
    \begin{tabular}{|c c || c | c | c | c  | c | }
    \hline
    \multicolumn{2}{ |c ||  }{\multirow{2}{*}{Img/Amp} } &
    \multicolumn{5}{ c| }{SNR (dB)}\\ \cline{3-7}
    \multicolumn{2}{ |c||  }{} & GA & LS & IRLS & PUMA & IRTV \\

	
    \hline\hline
    \multicolumn{1}{|c|}
    {\multirow{6}{*}{\rotatebox[origin=c]{90}{\textbf{Man}}}} & 
    4 & $\infbf$ & $\infbf$ & $\infbf$ & $\infbf$ & $\infbf$ \\ 
    \cline{2-7}
    
    \multicolumn{1}{|c|}{} & 
    5 & 26.29 & 32.31 & $\infbf$ & $\infbf$ & $\infbf$ \\ 
    \cline{2-7}
    
    \multicolumn{1}{|c|}{} & 
    6 & 10.12 & 24.55 & 30.47 & 29.43 & $\infbf$ \\ 
    \cline{2-7}
    
    \multicolumn{1}{|c|}{} & 
    7 & 11.70 & 18.14 & 24.23 & 27.46 & \textbf{29.22} \\ 
    \cline{2-7}
    
    \multicolumn{1}{|c|}{} & 
    8 & 8.75 & 10.17 & 21.89 & 12.08 & \textbf{25.26} \\ 
    \cline{2-7}
    
    \multicolumn{1}{|c|}{} & 
    9 & 6.95 & 6.75 & 12.69 & 8.36 & \textbf{13.23} \\ 
    \cline{1-7}
    
    
    \hline\hline
    \multicolumn{1}{|c|}
    {\multirow{6}{*}{\rotatebox[origin=c]{90}{\textbf{Lake}}}} & 
    4 & $\infbf$ & $\infbf$ & $\infbf$ & $\infbf$ & $\infbf$ \\ 
    \cline{2-7}
    
    \multicolumn{1}{|c|}{} & 
    5 & $\infbf$ & $\infbf$ & $\infbf$ & $\infbf$ & $\infbf$ \\ 
    \cline{2-7}
    
    \multicolumn{1}{|c|}{} & 
    6 & 40.10 & $\infbf$ & $\infbf$ & $\infbf$ & $\infbf$ \\ 
    \cline{2-7}
    
    \multicolumn{1}{|c|}{} & 
    7 & 26.62 & 25.21 & $\infbf$ & 41.44 & $\infbf$ \\ 
    \cline{2-7}
    
    \multicolumn{1}{|c|}{} & 
    8 & 15.99 & 13.84 & 18.02 & 17.69 & \textbf{18.10} \\ 
    \cline{2-7}
    
    \multicolumn{1}{|c|}{} & 
    9 & -0.88 & 2.84 & 15.91 & 15.33 & \textbf{16.52} \\ \cline{1-7}
    
    
    \hline\hline
    \multicolumn{1}{|c|}
    {\multirow{6}{*}{\rotatebox[origin=c]{90}{\textbf{Barbara}}}} & 
    4 & $\infbf$ & $\infbf$ & $\infbf$ & $\infbf$ & $\infbf$ \\ 
    \cline{2-7}
    
    \multicolumn{1}{|c|}{} & 
    5 & $\infbf$ & $\infbf$ & $\infbf$ & $\infbf$ & $\infbf$ \\ 
    \cline{2-7}
    
    \multicolumn{1}{|c|}{} & 
    6 & $\infbf$ & $\infbf$ & $\infbf$ & $\infbf$ & $\infbf$ \\ 
    \cline{2-7}
    
    \multicolumn{1}{|c|}{} & 
    7 & 43.54 & 40.53 & $\infbf$ & $\infbf$ & $\infbf$ \\ 
    \cline{2-7}
    
    \multicolumn{1}{|c|}{} & 
    8 & 35.60 & 30.07 & 44.70 & $\infbf$ & $\infbf$ \\ 
    \cline{2-7}
    
    \multicolumn{1}{|c|}{} & 
    9 & 31.01 & 22.01 & $\infbf$ & 45.72 & $\infbf$ \\ 
    \cline{1-7}
    
    \end{tabular}
    \egroup
\end{minipage}
\end{table*}

\begin{table*}[t]
\begin{center}
\footnotesize
\setlength{\tabcolsep}{2em}
  \caption{Phase unwrapping results for the noisy \emph{Man} image 
  of size $128 \times 128$ and amplitude $a = 6$.}\label{Tab:NoisyComparison}
  \def\arraystretch{1.3}
  \begin{tabular}{| l || c | c | c |}
  \hline
   Method / Input SNR & 16 dB & 18 dB & 20 dB \\ \hline\hline
   GA & 14.40 & 19.03 & 18.72 \\ \hline
   LS & 23.57 & 25.75 & 27.97 \\ \hline
   PUMA & 31.13 & 31.96 & 34.96 \\ \hline\hline
   IRTV $(\tau = 0)$ & 28.13 & 30.21 & 34.96 \\ \hline
   IRTV $(\tau = 10^{-3})$ & 28.52 & 31.97 & $\infbf$ \\ \hline
   IRTV $(\tau = 10^{-2})$ & \textbf{35.02} & \textbf{34.98} & $\infbf$ \\ \hline
   IRTV $(\tau = 10^{-1})$ & 18.86 & 24.56 & 28.46 \\
  \hline
  \end{tabular}
\end{center}
\end{table*}

\begin{figure*}[t]
\centering\includegraphics[width=12cm]{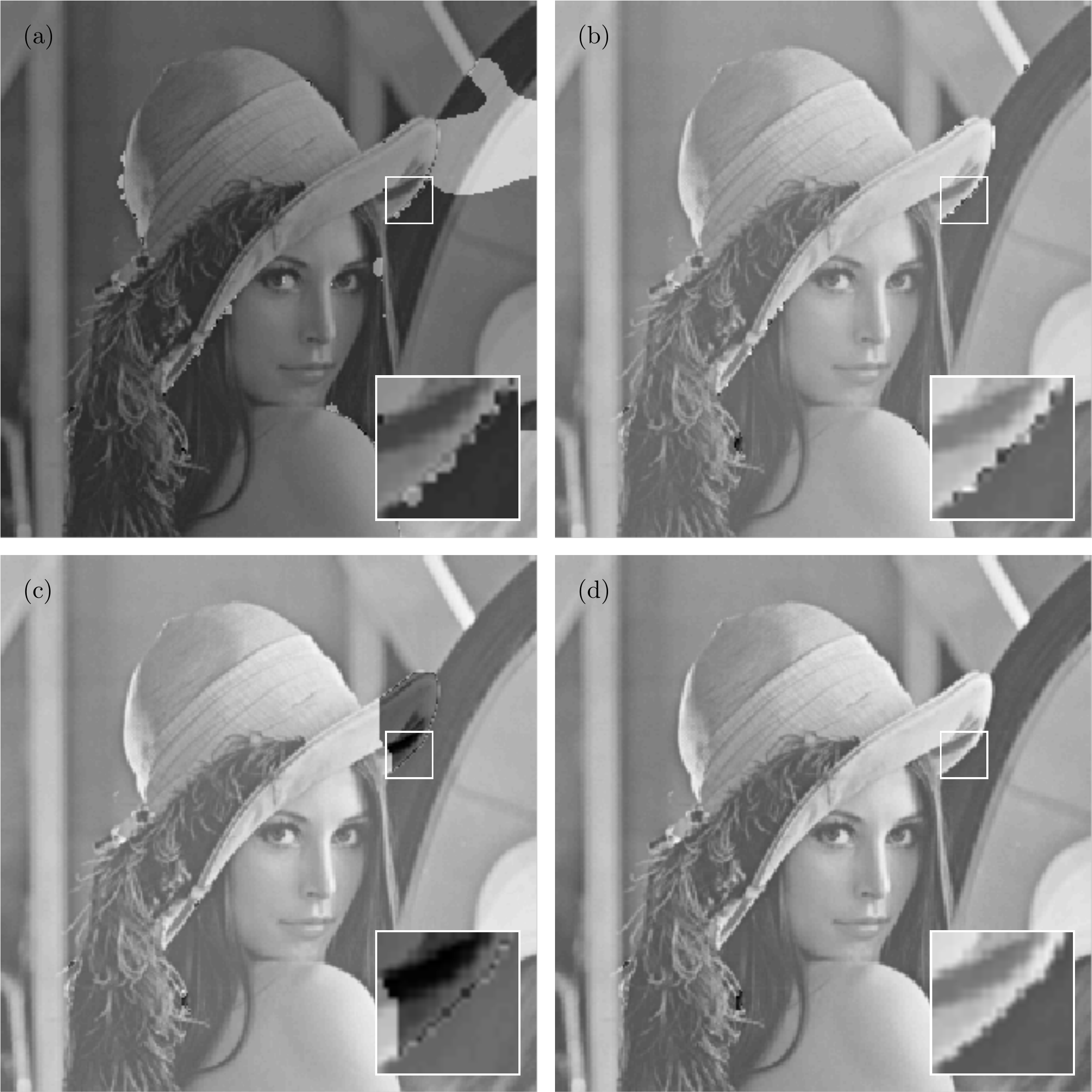}
\caption{Phase unwrapping results for \textit{Lena} image of size $256 \times 256$ when the amplitude is $a = 9$. (a) LS (SNR = 7.33 dB); (b) IRLS (SNR = 25.50 dB); (c) PUMA (SNR = 15.51 dB); (d) IRTV (SNR = 30.47 dB).}
\label{Fig:Lena}
\end{figure*}

\begin{figure*}[t]
\centering\includegraphics[width=12cm]{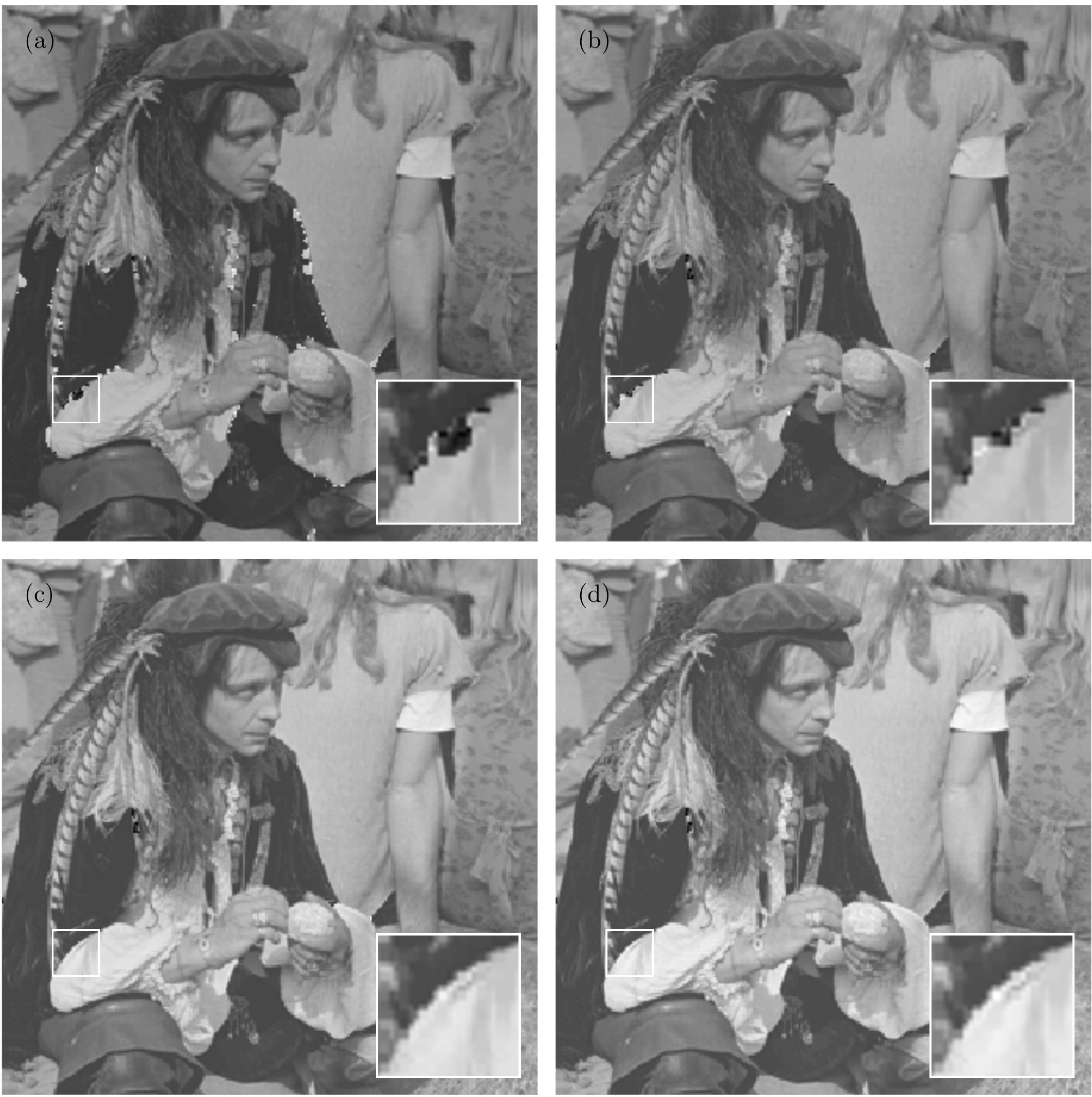}
\caption{Phase unwrapping results for \textit{Man} image of size $256 \times 256$ when the amplitude is $a = 7$. (a) LS (SNR = 18.14 dB); (b) IRLS (SNR = 24.23 dB); (c) PUMA (SNR = 27.46 dB); (d) IRTV (SNR = 29.22 dB).}
\label{Fig:Man}
\end{figure*}

\begin{figure*}[t]
\centering\includegraphics[width=16cm]{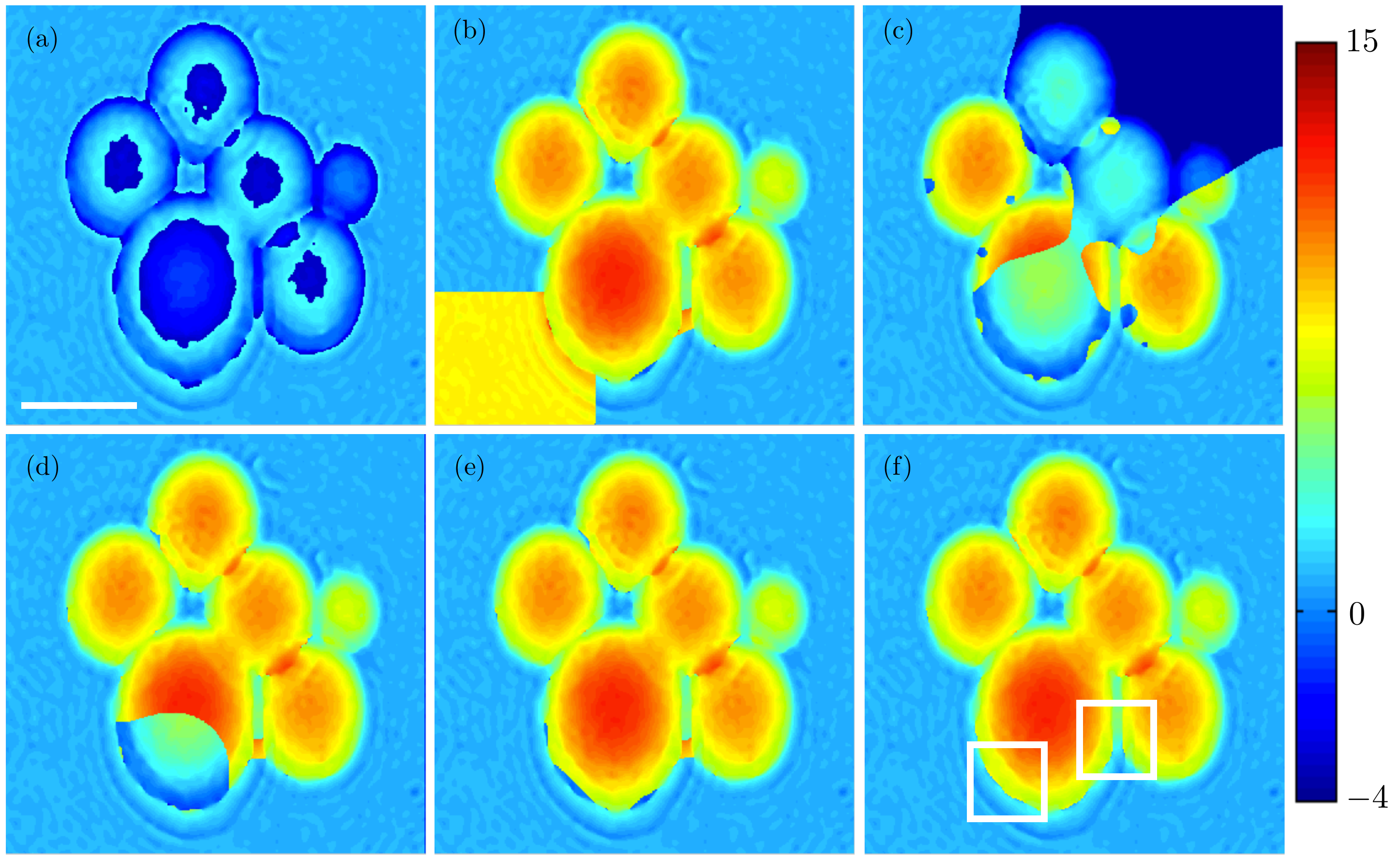}
\caption{Phase unwrapping results for \textit{Beads} image of 
size $512 \times 512$ obtained from the phase of the 
transmitted field. (a) wrapped phase; 
(b) GA; (c) LS; (d) IRLS; (e) PUMA; (f) IRTV. Scale bar, 5 $\mu$m.}
\label{Fig:Beads}
\end{figure*}

The step 2 of the algorithm requires the resolution of the non-smooth
optimization problem~\eqref{Eq:VariationalProblem}. 
We perform this minimization by designing an 
augmented-Lagrangian (AL) scheme~\cite{Nocedal.Wright2006}. Specifically,
we seek the critical points of the following AL
\begin{align*}
\lagrangian&(\phibf, \epsbf, \sbf) 
\defn \sum_{n = 1}^N w_n \|[\epsbf]_n\|_{\ell_2}+ \tau\regularizer(\phibf)  \\ 
&\hspace{1em}+ \sbf^{T} \left(\epsbf - \Dbf\phibf+\dbf\right) + \frac{\rho}{2}\|\epsbf - \Dbf\phibf+\dbf\|_{\ell_2}^2 \nonumber \\
&= \frac{\rho}{2} \left\|\epsbf - \Dbf\phibf + \dbf + \frac{\sbf}{\rho}\right\|_{\ell_2}^2 \\
&\hspace{1em}+ \sum_{n = 1}^N w_n \|[\epsbf]_n\|_{\ell_2} + \tau\regularizer(\phibf) - \frac{1}{2\rho} \|\sbf\|_{\ell_2}^2, \nonumber
\end{align*}
where $\dbf \defn \wrap(\Dbf\psibf)$ is the data vector, $\sbf \in \R^{N \times 2}$ is the dual variable 
that imposes the constraint $\epsbf = \Dbf\phibf-\dbf$,
and $\rho > 0$ is the quadratic penalty parameter. Traditionally, an AL scheme
solves the problem~\eqref{Eq:VariationalProblem} by alternating 
between a joint minimization step and an update step as
\begin{subequations}
\begin{align}
(\phibf^k, \epsbf^k) &\leftarrow \argmin_{\phibf \in \Phibf,\; \epsbf \in \R^{N \times 2}} \left\{\lagrangian(\phibf, \epsbf, \sbf^{k-1})\right\} \label{Eq:ALJoint}\\
\sbf^k &\leftarrow \sbf^{k-1} + \rho(\epsbf^k - \Dbf\phibf^k + \dbf ).
\end{align}
\end{subequations}
However, the joint minimization step~\eqref{Eq:ALJoint} can be computationally intensive.
To circumvent this problem, we separate~\eqref{Eq:ALJoint} into a succession
of simpler steps. This form of separation is commonly known as
alternating direction method of multipliers (ADMM)~\cite{Boyd.etal2011}
and can be described as follows
\begin{subequations}
\label{Eq:ADMM}
\begin{align}
\phibf^k &\leftarrow \argmin_{\phibf \in \Phibf} \left\{\lagrangian(\phibf, \epsbf^{k-1}, \sbf^{k-1})\right\} \label{Eq:OptPhi}\\
\epsbf^k &\leftarrow \argmin_{\epsbf \in \R^{N \times 2}} \left\{\lagrangian(\phibf^k, \epsbf, \sbf^{k-1})\right\} \label{Eq:OptEps}\\
\sbf^k &\leftarrow \sbf^{k-1} + \rho(\epsbf^k - \Dbf\phibf^k + \dbf). \label{Eq:OptDual}
\end{align}
\end{subequations}
By ignoring the terms that do not depend on $\phibf$, the step~\eqref{Eq:OptPhi} can be expressed as
\begin{equation}
\label{Eq:HSMinimization}
\phibf^k \leftarrow \argmin_{\phibf \in \Phibf} \left\{\frac{1}{2} \left\|\Dbf\phibf - \zbf^{k-1}\right\|_{\ell_2}^2 + \frac{\tau}{\rho}\regularizer(\phibf)\right\},
\end{equation}
with $\zbf^{k-1} \defn \epsbf^{k-1} + \dbf + \sbf^{k-1}/\rho$. This step corresponds to a classical Hessian Schatten--regularized linear inverse problem. 
The solution of this problem can be efficiently solved with the publicly available software that
has been described in~\cite{Lefkimmiatis.etal2013}.
Similarly, the step~\eqref{Eq:OptEps} can be simplified as follows
\begin{equation*}
\epsbf^k \leftarrow \argmin_{\epsbf \in \R^{N \times 2}} \left\{\frac{\rho}{2} \left\|\epsbf - \ybf^{k}\right\|_{\ell_2}^2 + \sum_{n = 1}^N w_n \|[\epsbf]_n\|_{\ell_2}\right\},
\end{equation*}
with $\ybf^{k} \defn \Dbf\phibf^k - \dbf - \sbf^{k-1}/\rho$.
This step is solved directly by component-wise application of the following shrinkage function
\begin{align*}
\shrinkage(\ybf; \tau) 
&\defn \argmin_{\xbf \in \R^2} \left\{\frac{1}{2}\|\xbf - \ybf\|_{\ell_2}^2 + \tau\|\xbf\|_{\ell_2}\right\} \\
&= \max\left(\|\ybf\|_{\ell_2} - \tau, 0\right) \frac{\ybf}{\|\ybf\|_{\ell_2}}.
\end{align*}
Thus, we can express~\eqref{Eq:OptEps} as
\begin{equation*}
[\epsbf^k]_n \leftarrow \shrinkage\left(\left[\Dbf\phibf^k - \dbf - \sbf^{k-1}/\rho\right]_n; w_n/\rho\right),
\end{equation*}
for every $n = 1, \dots, N$.

While the theoretical
convergence of our algorithm requires the full convergence of ADMM 
inner iterations~\eqref{Eq:ADMM},
in practice, we found that, by using a sufficiently high number of iterations 
$k_{\textrm{max}}$ with an additional stopping criterion, 
our algorithm achieves excellent results 
as illustrated in Sec.~\ref{Sec:Experiments}. In particular,
we implemented the standard criterion suggested
by Boyd \emph{et al.}~\cite{Boyd.etal2011}, where ADMM is
stopped when then primal and dual residuals are small
\begin{subequations}
\label{Eq:InnerCriteria}
\begin{align}
&\|\epsbf^k - \Dbf\phibf^k - \dbf\|_{\ell_2} \leq \delta_{\textrm{inner}} \\ 
&\|\rho\Dbf(\phibf^k - \phibf^{k-1})\|_{\ell_2} \leq \delta_{\textrm{inner}},
\end{align}
\end{subequations}
where the constant $\delta_{\textrm{inner}} > 0 $ controls the desired inner tolerance level.
In all our experiments, we set $\delta_{\textrm{inner}} = 10^{-2}$ and $k_{\textrm{max}} = 2000$.

For outer iterations, our algorithm relies on a separate number of $t_{\textrm{max}}$
iterations with a distinct stopping criterion based on
measuring the relative change of the solution in two successive iterations as
\begin{equation}
\label{Eq:OuterCriteria}
\frac{\|\phibf^t-\phibf^{t-1}\|_{\ell_2}}{\|\phibf^{t-1}\|_{\ell_2}} \leq \delta_{\textrm{outer}},
\end{equation}
where $\delta_{\textrm{outer}} > 0 $ controls the desired outer tolerance level. 
In the experiments, these constants were set to $t_{\textrm{max}} = 10$
and $\delta_{\textrm{outer}} = 10^{-2}$.

It is important to note that the solution of our iterative method
is not consistent in the sense that the rewrapped phase
$\wrap(\phibfhat)$ is not necessarily equal to 
the measured phase $\psibf$. 
The possible inconsistency of our solution comes from the
fact that we are using a continuous optimization for solving
an inherently discrete optimization problem (i.e., addition
and subtraction of integer multiples of $2\pi$). Accordingly,
any method relying on continuous optimization such as
LS, IRLS, or the method proposed here, may result in
an inconsistent solution. Path-following methods
such as Goldstein's algorithm or discrete optimization algorithms
such as PUMA always return consistent solutions.
Consistency, however, is easily achieved with 
a single post-processing step 
that was proposed by Pritt~\cite{Pritt1997} as
\begin{equation}
\label{Eq:CongruencyOperator}
\phibfhat^\prime \leftarrow \phibfhat + \wrap(\psibf - \phibfhat),
\end{equation}
where $\phibfhat^\prime$ is consistent with the wrapped phase $\psibf$.
In the sequel, we perform a single application of the 
operator~\eqref{Eq:CongruencyOperator} to the outputs returned by
the continuous optimization algorithms to make their solutions
consistent with the measurements.

To conclude, we described a method for minimizing 
the proposed objective functional. While this forces us to revert to a more costly
iterative scheme (instead of finding the solution directly), it allows us to 
obtain a variational formulation that incorporates the most efficient
ideas that have appeared for phrase unwrapping and the stabilization of 
ill-posed inverse problems. The optimization itself is performed iteratively 
by relying on ADMM to reduce the problem to a 
succession of straightforward operations. The final computational time required to 
unwrap a given image depends on the total number of iterations, 
which in turn depends on the severity of wrapping. For example, it took us
about 3 minutes to unwrap the Gaussian signal in Fig.~$\ref{Fig:Gaussian}$
with a MATLAB implementation of our algorithm running on an iMac using
a 4 GHz Intel Core i7 processor. In Section~\ref{Sec:Experiments}, we illustrate
that the improvement in reconstruction quality can be
rather substantial. We thus believe that the method should be of interest to practitioners
that rely on quantitative phase evaluation.

\section{Experiments} \label{Sec:Experiments}

Based on the above developments, we report the results of our 
phase unwrapping method in simulated and practical configurations.
In particular, we compare the results of our approach, 
which we shall denote IRTV,
against those obtained by using four alternative methods;
namely, Golstein's algorithm (GA)~\cite{Goldstein.etal1988}, 
least-squares (LS)~\cite{Ghiglia.Romero1994}, iterative
reweighed LS approach (IRLS)
with data-dependent weights that
approximate the $\ell_0$-norm penalty~\cite{Ghiglia.Romero1996}, 
and PUMA~\cite{Bioucas-Dias.Valadao2007}. 
An implementation of PUMA is freely available at \url{http://www.lx.it.pt/ ~bioucas/code.htm}.
Mirroring examples provided by the authors of PUMA, 
we use it with a nonconvex quantized
potential of exponent $p = 0.5$ where quadratic region threshold
is set to $0.5$. As suggested in the PUMA code, we use cliques of
higher order by considering $4$ displacement vectors $(1, 0), (0.1),
(1, 1)$ and $(-1, 1)$.

In the sequel, a first set of experiments 
with simulated phase wrapping evaluates the algorithms 
quantitatively by using the true phase $\phibf$
for comparison. The experiments that follow allow to determine 
the relevance of our approach
on real data in the context of tomographic phase microscopy.

\subsection{Synthetic Data}

In this set of experiments, we use a set of $6$ grayscale 
$256 \times 256$ shown in Fig.~\ref{Fig:ImageDatabase} from the USC-SIPI database at 
\url{http://sipi.usc.edu/database/}.
After normalization of its amplitude to $[0, a]$, 
with $a \in [4, 5, 6, 7, 8, 9]$, each image is used to generate 
a distinct wrapped phase image according to our observation model~\eqref{Eq:Model}.
As image amplitude $a$ increases, the edges in
the true phase $\phibf$ become sharper,
thus making the unwrapping task increasingly more difficult.

Given the wrapped phase image $\psibf$, our goal is to determine how 
accurately the true phase $\phibf$ can be reconstructed by the 
standard and proposed methods. As reconstruction parameters,
our algorithm uses $\tau = 10^{-2}$ and $\epsilon_{\textrm{max}} = 1/\epsilon_{\textrm{min}} = 10$ in all synthetic experiments. As mentioned before, we
set the number of outer and inner iterations to 
$t_{\textrm{max}} = 10$ 
and $k_{\textrm{max}} = 2000$, respectively, with
additional stopping criteria~\eqref{Eq:InnerCriteria} and~\eqref{Eq:OuterCriteria}.
We also use $10$ iterations for finding the solution of 
Hessian Schatten-regularized inverse problem~\eqref{Eq:HSMinimization}.
In principle, the positive scalar $\rho$ of ADMM can either be fixed to 
some predetermined value 
or adapted according to the distance to the 
constraint via the scheme described 
in~\cite{Boyd.etal2011}. In all our experiments, 
we used adaptive $\rho$; however,
we found that in practice fixed values of $\rho = 0.1$ 
and $\rho = 1$ work equally well. Iterative algorithms IRLS and
IRTV were initialized with the LS solution.
Upon convergence,
the solutions of all the variational algorithms
were made consistent with a single application of the
operator~\eqref{Eq:CongruencyOperator}.

In Table~\ref{Tab:Comparison}, we report the signal-to-noise ratio
$\textrm{SNR (dB)} \defn 10\log_{10}\left(\|\phibf\|_{\ell_2}^2/\|\phibf-\phibfhat\|_{\ell_2}^2\right)$
of the unwrapped phase for all the methods considered. When the SNR is 
more that $300$ dB, we consider the unwrapping
to be exact and set the corresponding value in the table to infinity. 
As can be seen from the table, our method, which is labeled IRTV, consistently provides
better reconstruction results for nearly all images and all amplitude
values.  Specifically, our method successfully unwraps  
of \textit{Cameraman}, \textit{Lena}, and \textit{Man} phase images
for which all other standard methods fail. Note that for \textit{Cameraman}
at amplitudes 5, 8, and 9 LS yields unwrapped phase images 
of higher quality. While for $a = 5$ the difference between LS and IRTV 
is modest (less that 0.15 dB), 
for $a = 8$ and $a = 9$ all the methods completely fail at unwrapping (SNRs below 3 dB).

Beyond the SNR comparisons, the effectiveness of the proposed 
method can also be visually appreciated by inspecting the 
more difficult scenarios of \textit{Lena} and \textit{Man} 
images presented in Fig.~\ref{Fig:Lena} and Fig.~\ref{Fig:Man}. From these 
examples we can verify our initial claim: our phase unwrapping
method recovers oriented sharp edges more accurately than other algorithms
thus providing an overall boost in performance.

In order to illustrate the robustness of IRTV to noise,
we now consider a simple scenario where the unwrapped phase 
$\phibf$ corresponds to a noisy version of the true phase $\varphibf$.
Specifically, we consider the additive noise model $\phibf = \varphibf + \ebf$,
where $\ebf$ is additive white Gaussian noise (AWGN).
Given the wrapped image $\psibf = \wrap(\phibf) = \wrap(\varphibf + \ebf)$, 
we would like to determine how accurately one can recover $\phibf$
in the presence of high levels of noise.
Noisy phase images are typically more difficult to unwrap due to additional
discontinuities that appear across the whole image.
Note that robustness to noise is different from denoising, which
implies that we do not attempt to reduce the level of noise during 
the unwrapping process. When unwrapping is successful, one 
can then denoise the unwrapped image with any state-of-the-art 
image denoising algorithm suitable to the noise at hand.

In Table~\ref{Tab:NoisyComparison}, we report the SNR of the
the estimate $\phibfhat$ with respect to the unwrapped phase $\phibf$
for GA, LS, PUMA, and IRTV algorithms. The true phase
$\varphibf$ is the \emph{Man} 
image of amplitude $a = 6$ and dimension $128 \times 128$. 
The phase image $\phibf$ is obtained by adding 
AWGN of variance corresponding
to 16, 18, and 20 dB of $\textrm{input SNR (dB)} \defn 10\log_{10}\left(\|\varphibf\|_{\ell_2}^2/\|\phibf-\varphibf\|_{\ell_2}^2\right)$. The table presents the median SNR for
$10$ independent realizations of the noise. In this example, the average 
running time of our MATLAB implementation of IRTV 
was about 7.5 minutes on our iMac using a 4 GHz Intel Core i7 processor. 
Additionally, the table presents
the results of IRTV with different values of regularization parameter $\tau$.
Specifically, we report the results for $\tau \in [0, 10^{-3}, 10^{-2}, 10^{-1}]$.
Higher levels of $\tau$ imply stronger regularization during the reconstruction.
The results in the table illustrate the advantage of using 
the Schatten norm of the Hessian to complement our data-fidelity term, i.e., one obtains a 
significant boost in unwrapping performance when $\tau > 0$. Moreover, the influence of
$\tau$ grows as the level of noise increases from 20 to 16 dB of input SNR. 
One must however note that similar to other regularization schemes, 
there is no theoretically optimal way of setting 
$\tau$ and its optimal value might depend on the image, amount of wrapping, and noise.
Our simulations indicate that the optimal value of $\tau$ lies in the range $[10^{-3}, 10^{-1}]$
for the configurations considered.

\subsection{Real Data}

In this second experimental part, we consider phase images 
that are acquired practically from distinct physical objects.
In particular, we consider the setup for tomographic phase microscopy~\cite{Choi.etal2007}, which is a promising quantitative phase 
imaging technique. It is based on the principle that,
for near-plane wave illumination of a sample, 
the phase of the transmitted field 
can be well approximated as the integral of the refractive
index along the path of beam propagation. However, 
for the approximation to hold, the phases extracted from
the transmitted fields must be first unwrapped, which significantly
limits the applications of the technique for imaging objects that are
off-focus, large, or have high index contrast.
Once unwrapped, the phase image can simply be interpreted as the 
projection of refractive index, analogous to the projection 
of absorption in X-ray tomography.

To evaluate our phase unwrapping algorithm, we measured refractive
index tomograms of 6 polystyrene spheres (catalog no.\ 17135, Polysciences,  
refractive index $n=1.602$ at 561 nm) 
immersed in oil with a lower refractive index of 1.516. The
wrapped phase image of size $512 \times 512$ is extracted from the 
transmitted field at angle $39.02^\textrm{o}$ with respect to vertical
axis. This phase data is difficult to unwrap due of numerous visible phase 
discontinuities that appear along the borders of the beads.

The parameters of our IRTV method were chosen
as in the synthetic experiments above with the exception
of the regularization parameter that was set to 
$\tau = 10^{-1}$.
Also as above, the reconstructed phase images 
were made consistent by using the
operator~\eqref{Eq:CongruencyOperator}. 

The results in Fig.~\ref{Fig:Beads} illustrate the effectiveness
of our method in unwrapping the phase even in the most difficult
regions of the image that contain strong phase discontinuities.
Specificaly, our method is the only one that was able to 
accurately unwrap the region between the two beads at the bottom
of the image (see highlights in the figure).

\section{Conclusion} \label{Sec:Conclusion}

We have devised an algorithm for two-dimensional phase unwrapping 
that is based on an isotropic problem formulation. Based on 
suitable regularity assumptions, our technique has allowed 
to unwrap various phase images satisfactorily, including in the case 
where the phase contained significant amount of discontinuities. 
Compared to the standard techniques, the proposed method 
preserves edges of arbitrary orientations in the solution 
and effectively mitigates noise in practical 
configurations. From a general perspective, the obtained 
results further illustrate the interest of inverse-problem 
approaches for phase unwrapping. 

\bibliographystyle{alpha}

\begin{thebibliography}{BDKAE08}

\bibitem[BDKAE08]{Bioucas-Dias.etal2008}
J.~M.~Bioucas-Dias, V.~Katkovnik, J.~Astola, and K.~Egiazarian.
\newblock Absolute phase estimation: adaptive local denoising and global
  unwrapping.
\newblock {\em Appl. Opt.}, 47(29):5358--5369, 2008.

\bibitem[BDV07]{Bioucas-Dias.Valadao2007}
J.~M. Bioucas-Dias and G.~Valad{\~a}o.
\newblock Phase unwrapping via graph cuts.
\newblock {\em IEEE Trans. Image Process.}, 16(3):698--709, March 2007.

\bibitem[BPC{\etalchar{+}}11]{Boyd.etal2011}
S.~Boyd, N.~Parikh, E.~Chu, B.~Peleato, and J.~Eckstein.
\newblock Distributed optimization and statistical learning via the alternating
  direction method of multipliers.
\newblock {\em Foundations and Trends in Machine Learning}, 3(1):1--122, 2011.

\bibitem[BT09]{Beck.Teboulle2009a}
A.~Beck and M.~Teboulle.
\newblock Fast gradient-based algorithm for constrained total variation image
  denoising and deblurring problems.
\newblock {\em IEEE Trans. Image Process.}, 18(11):2419--2434, November 2009.

\bibitem[CFYB{\etalchar{+}}07]{Choi.etal2007}
W.~Choi, C.~Fang-Yen, K.~Badizadegan, S.~Oh, N.~Lue, R.~R. Dasari, and M.~S.
  Feld.
\newblock Tomographic phase microscopy.
\newblock {\em Nat. Methods}, 4(9):717--719, September 2007.

\bibitem[CWB08]{Candes.etal2008}
E.~J. Cand{\`e}s, M.~B. Wakin, and S.~P. Boyd.
\newblock Enhancing sparsity by reweighted $\ell_1$ minimization.
\newblock {\em J. of Fourier Anal. Appl.}, 14(5--6):877--905, October 2008.

\bibitem[Fri77]{Fried1977}
D.~L. Fried.
\newblock Least-square fitting a wave-front distortion estimate to an array of
  phase-difference measurements.
\newblock {\em J. Opt. Soc. Am.}, 67(3):370--375, 1977.

\bibitem[GJ14]{Gonzalez.Jacques2014}
A.~Gonz{\'{a}}lez and L.~Jacques.
\newblock Robust phase unwrapping by convex optimization.
\newblock In {\em Proc. IEEE Int. Conf. Image Process ({ICIP}'14)}, Paris,
  France, October 27--30, 2014.
\newblock {arXiv:1407.8040 [math.OC]}.

\bibitem[GP98]{Ghiglia.Pritt1998}
D.~C. Ghiglia and M.~D. Pritt.
\newblock {\em Two-Dimensional Phase Unwrapping: Theory, Algorithms, and
  Software}.
\newblock John Willey \& Sons, 1998.

\bibitem[GR94]{Ghiglia.Romero1994}
D.~C. Ghiglia and L.~A. Romero.
\newblock Robust two-dimensional weighted and unweighted phase unwrapping that
  uses fast transforms and iterative methods.
\newblock {\em J. Opt. Soc. Am.}, 11(1):107--117, 1994.

\bibitem[GR96]{Ghiglia.Romero1996}
D.~C. Ghiglia and L.~A. Romero.
\newblock Minimum {$L^p$}-norm two-dimensional phase unwrapping.
\newblock {\em J. Opt. Soc. Am. A}, 13(10):1999--2013, October 1996.

\bibitem[GZW88]{Goldstein.etal1988}
R.~M. Goldstein, H.~A. Zebker, and C.~L. Werner.
\newblock Sattelite radar interferometry: {T}wo-dimensional phase unwrapping.
\newblock {\em Radio Sci.}, 23(4):713--720, July--August 1988.

\bibitem[HBLG02]{Herraez.etal2002}
M.~Arevallilo Herr{\'{a}}ez, D.~R. Burton, M.~J. Lalor, and
  M.~A. Gdeisat.
\newblock Fast two-dimensional phase-unwrapping algorithm based on sorting by
  reliability following a noncontinuous path.
\newblock {\em Appl. Opt.}, 41(35):7437--7444, December 2002.

\bibitem[HTZ{\etalchar{+}}12]{Huang.etal2012}
{H.~Y.~H.} Huang, L.~Tian, Z.~Zhang, Y.~Liu, Z.~Chen, and G.~Barbastathis.
\newblock Path-independent phase unwrapping using phase gradient and
  total-variation ({TV}) denoising.
\newblock {\em Opt. Express}, 20(13):14075--14089, June 2012.

\bibitem[Hun79]{Hunt1979}
B.~R. Hunt.
\newblock Matrix formulation of the reconstruction of phase values from phase
  differences.
\newblock {\em J. Opt. Soc. Am.}, 69(3):393--399, March 1979.

\bibitem[Ito82]{Itoh1982}
K.~Itoh.
\newblock Analysis of the phase unwrapping problem.
\newblock {\em Appl. Opt.}, 21(14):2470, July 1982.

\bibitem[LWU13]{Lefkimmiatis.etal2013}
S.~Lefkimmiatis, J.~P.~Ward, , and M.~Unser.
\newblock Hessian {S}chatten-norm regularization for linear inverse problems.
\newblock {\em IEEE Trans. Image Process.}, 22(5):1873--1888, May 2013.

\bibitem[MKCG13]{Mei.etal2013}
J.~Mei, A.~Kirmani, A.~Cola{\c{c}}o, and V.~K Goyal.
\newblock Phase unwrapping and denoising for time-of-flight imaging using
  generalized approximate message passing.
\newblock In {\em Proc. IEEE Int. Conf. Image Process ({ICIP}'13)}, pages
  364--368, Melbourne, VIC, Australia, September 15--18, 2013.

\bibitem[MR95]{Marroquin.Rivera1995}
J.~L. Marroquin and M.~Rivera.
\newblock Quadratic regularization functionals for phase unwrapping.
\newblock {\em J. Opt. Soc. Am.}, 12(11):2393--2400, 1995.

\bibitem[NW06]{Nocedal.Wright2006}
J.~Nocedal and S.~J. Wright.
\newblock {\em Numerical Optimization}.
\newblock Springer, 2 edition, 2006.

\bibitem[Pri97]{Pritt1997}
M.~D. Pritt.
\newblock Congruence in least-squares phase unwrapping.
\newblock In {\em 1997 IEEE International Geoscience and Remote Sensing
  Symposium (IGARSS)}, volume~2, pages 875--877, Singapore, August 03--08,
  1997.

\bibitem[RHJ{\etalchar{+}}00]{Rosen.etal2000}
P.~A. Rosen, S.~Hensley, I.~R. Joughin, F.~K. Li, S.~N.~Madsen,
  E.~Rodr{\'{i}}guez, and R.~M.~Goldstein.
\newblock Synthetic aperture radar interferometry.
\newblock {\em Proc. IEEE}, 88(3):333--382, March 2000.

\bibitem[RM04]{Rivera.Marroquin2004}
M.~Rivera and J.~L. Marroquin.
\newblock Half-quadratic cost functions for phase unwrapping.
\newblock {\em Opt. Lett.}, 29(5):504--506, 2004.

\bibitem[SNPG95]{Song.etal1995}
S.~M.-H. Song, S.~Napel, N.~J. Pelc, and G.~H. Glover.
\newblock Phase unwrapping of {MR} phase images using {P}oisson equation.
\newblock {\em IEEE Trans. Image Process.}, 4(5):667--676, May 1995.

\bibitem[TT88]{Takajo.Takahashi1988}
H.~Takajo and T.~Takahashi.
\newblock Least-squares phase estimation from the phase difference.
\newblock {\em J. Opt. Soc. Am. A}, 5(3):416--425, 1988.

\bibitem[VBD09]{Valadao.Biocas-Dias2009}
G.~Valad{\~a}o and J.~Biocas-Dias.
\newblock {CAPE}: combinatorial absolute phase estimation.
\newblock {\em J. Opt. Soc. Am. A}, 26(9):2093--2106, September 2009.

\bibitem[Yin06]{Ying2006}
L.~Ying.
\newblock {\em Wiley Encyclopedia of Biomedical Engineering}, chapter Phase
  unwrapping.
\newblock John Wiley \& Sons, 2006.

\bibitem[YLJ{\etalchar{+}}06]{Ying.etal2006}
L.~Ying, Z.-P. Liang, D.~C.~Munson Jr., R.~Koetter, and B.~J. Frey.
\newblock Unwrapping of {MR} phase images using a {M}arkov random field model.
\newblock {\em IEEE Trans. Med. Imag.}, 25(1):128--136, January 2006.

\end{thebibliography}

\newcommand{\etalchar}[1]{$^{#1}$}

\end{document}